\documentclass[aps,pra,twocolumn,superscriptaddress,showpacs]{revtex4-1}
\def\be{\begin{equation}}
\def\ee{\end{equation}}
\def\bea{\begin{eqnarray}}          
\def\eea{\end{eqnarray}}
\def\bi{\begin{itemize}}
\def\ei{\end{itemize}}

\usepackage{graphicx}

\begin{document}

\title{ 
A ring of BEC pools as a trap for persistent flow
}

\author{Jacek Dziarmaga} 
\affiliation{Instytut Fizyki Uniwersytetu Jagiello\'nskiego, 
             %and Centre for Complex Systems Research
             ul. Reymonta 4, 30-059 Krak\'ow, Poland}

\author{Marek Tylutki}
\affiliation{Instytut Fizyki Uniwersytetu Jagiello\'nskiego, 
             %and Centre for Complex Systems Research
             ul. Reymonta 4, 30-059 Krak\'ow, Poland}

\author{Wojciech H. Zurek}
\affiliation{Theory Division, Los Alamos National Laboratory, Los Alamos, NM 87545, USA}

\date{ July 25, 2011 }

\begin{abstract}
Mott insulator - superfluid transition in a periodic lattice of Josephson junctions can be driven by tunneling rate increase. Resulting winding numbers $W$ of the condensate wavefunction decrease with increasing quench time in accord with the Kibble-Zurek mechanism (KZM). However, in very slow quenches Bose-Hubbard dynamics rearranges wavefunction phase so that its random walk cools, $\overline{W^2}$ decreases and eventually the wavefunction becomes too cold to overcome potential barriers separating different $W$. Thus, in contrast with KZM, in very slow quenches $\overline{W^2}$ is set by random walk with ``critical'' step size, independently of $\tau_Q$. As our study requires use of the truncated Wigner approximation (TWA) over relatively long time intervals, we investigate validity of TWA by comparing its predictions with exact calculations for suitably small quantum systems.
\end{abstract}

\maketitle

%%%%%%%%%%%%%%%%%%%%%%%%%%%%%%%%%%%%%%%%%%%%%%%%%%%%%%%%%%%%%%%%%%%%%%%%%%%%%%%

\section{Introduction} 

%%%%%%%%%%%%%%%%%%%%%%%%%%%%%%%%%%%%%%%%%%%%%%%%%%%%%%%%%%%%%%%%%%%%%%%%%%%%%%%%%%%%%%

\begin{figure}[t]
\includegraphics[width=0.9\columnwidth,clip=true]{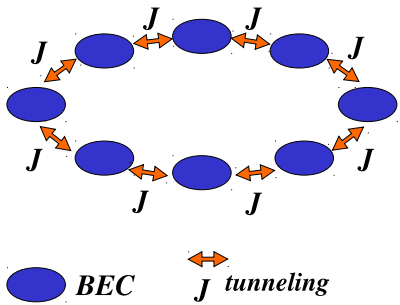}
\caption{ 
An initial ring of isolated Bose-Einstein condensates, each with a definite large number of particles $n$ and indefinite phase, becomes phase-correlated by slowly turning on the tunneling between the condensates, see Refs. \cite{Mercedes,Boshier}.
}
\label{FigScheme}
\end{figure}

Gapless quantum critical points are a serious obstacle for quantum simulation with ultracold atomic gases or ion traps, where one would like to prepare a simple ground state of a simple initial Hamiltonian and then drive the Hamiltonian adiabatically to an interesting final ground state. This general observation has been recently substantiated by a more quantitative theory \cite{QuantumKZ} (see \cite{review} for reviews) which is a quantum generalization of the classical Kibble-Zurek mechanism (KZM) \cite{KZ}. The theory predicts that density of excitations (or excitation energy) decays with (usually a fractional) power of quench rate $1/\tau_Q$. Experiments related to the quantum theory, although in the sudden quench limit $\tau_Q\to0$, were made in Refs. \cite{ferro,Mercedes}. In one of them \cite{Mercedes} a ring of $N=3$ isolated Bose-Einstein condensates (BEC's) was connected, see Fig. \ref{FigScheme}. Random initial phases of BEC's often result in a non-zero vorticity trapped in the final state. A similar ring with a bigger $N$ can be prepared by painting a time-dependent potential, as is demonstrated in Ref. \cite{Boshier}.  

We use the truncated Wigner method \cite{TW,TWAAnatoli} in the Bose-Hubbard model to simulate
gradually connecting a ring of condensates.  For sufficiently slow quenches we find that typical winding numbers trapped in the connected ring do not depend on how slowly BECs are connected. The process responsible for the final trapped winding numbers can be thought of as frustration -- trapping of the phase of the cooling, equilibrated condensate wavefunction in one of the minima corresponding to integer $W$. 

In contrast to KZM (where winding number $W$ depends on the time $\hat t$ when the critical slowing down ceased to suppress phase-ordering dynamics \cite{Meisner}), for sufficiently slow quenches the condensate phase evolves ergodically, exploring different potential minima. 
Therefore, instantaneous half-width of a Gaussian distribution of winding numbers is  determined by the step size of the random walk and proportional to the square root of the number of sites -- its length. Trapping of this equilibrium winding occurs as the size of random walk steps -- its ``temperature"  -- becomes too small to get over potential barriers. Final $W$ are set by the critical step size $\sigma_c$ -- the least step size that allows for jumps -- and yield $W^2\approx\sigma_c^2N$.

Thus, sufficiently slow quenches lose memory of phase dispersion and of $W$ at $\hat t$, which in faster quenches was shown to lead to $W^2$ consistent with KZM \cite{Meisner}. In the units (set by the dimensionless Hamiltonian immediately below) this (wide!) border turns out to be near $\tau_Q\sim 10^3$. 

This paper starts with a brief dicussion of the Bose-Hubbard model (the subject of our study) in Section \ref{SectionBHM} and of the truncated Wigner approximation (our principal tool) in Section \ref{SectionTWA}. Before presenting our results, we digress to explore the validity of TWA by comparing its predictions to the exact calculations of small quantum systems. This is done in the extensive Appendix. After we have introduced the model and validated the method we discuss results of our simulations and provide an analytic understanding and a heuristic picture of the winding number trapping in very slow quenches in Sections \ref{SectionTWA}-\ref{SectionPhonons}. We conclude in Section \ref{SectionConclusion}.

%%%%%%%%%%%%%%%%%%%%%%%%%%%%%%%%%%%%%%%%%%%%%%%%%%%%%%%%%%%%%%%%%%%%%%%%%%%%
%%%%%%%%%%%%%%%%%%%%%%%%%%%%%%%%%%%%%%%%%%%%%%%%%%%%%%%%%%%%%%%%%%%%%%%%%%%%

%%%%%%%%%%%%%%%%%%%%%%%%%%%%%%%%%%%%%%%%%%%%%%%%%%%%%%%%%%%%%%%%%%%%%%%%%%%%%%%%%%%%%%

\section{ Bose-Hubbard model }\label{SectionBHM}

%%%%%%%%%%%%%%%%%%%%%%%%%%%%%%%%%%%%%%%%%%%%%%%%%%%%%%%%%%%%%%%%%%%%%%%%%%%%%%%%%%%%%%

The model describes spinless cold bosonic atoms in an optical lattice \cite{Kasevich,Greiner}. In dimensionless variables, its Hamiltonian reads
\be
H = -J \sum_{s=1}^N \left( a_{s+1}^\dag a_s + {\rm h. c.} \right)
  + \frac{1}{2n} \sum_{s=1}^N a_s^\dag a_s^\dag a_s a_s ~
\label{H}
\ee
with periodic boundary conditions. Here $N$ is a number of lattice sites and $n$ is average number of atoms per site. For an integer $n$, there is a quantum phase transition from the Mott insulator to superfluid at $J_{cr}\simeq n^{-2}$ \cite{KT}. 

We drive the system by a linear quench 
\be
J(t)~=~{t}/{\tau_Q}~,
\label{Jt}
\ee
starting in the Mott ground state at $J=0$, 
\be
|n,n,n,\dots,n\rangle~,
\label{Mott}
\ee
with the same large number of particles $n$ at each site, and ending in the Josephson regime
\be
J~\ll~1
\ee 
where the interactions still dominate over the hopping.    

%%%%%%%%%%%%%%%%%%%%%%%%%%%%%%%%%%%%%%%%%%%%%%%%%%%%%%%%%%%%%%%%%%%%%%%%%%%%%

\section{ Truncated Wigner approximation (TWA) }\label{SectionTWA}

%%%%%%%%%%%%%%%%%%%%%%%%%%%%%%%%%%%%%%%%%%%%%%%%%%%%%%%%%%%%%%%%%%%%%%%%%%%%%%%%%%%%%%

For large density $n$ we replace annihilation operators $a_s$ by a complex field $\phi_s$, $a_s\approx\sqrt{n}\phi_s$, normalized as $\sum_{s=1}^N |\phi_s|^2=N$ and evolving with the Gross-Pitaevskii equation (GPE)
\be
i\frac{d\phi_s}{dt} = -J\left(\phi_{s+1}-2\phi_s+\phi_{s-1}\right) + 
\left(|\phi_s|^2-1\right)\phi_s ~.
\label{GPE}
\ee
Quantum expectation values are estimated by averages over stochastic realizations of $\phi_s$. Each realization has different random initial conditions coming from a Wigner distribution of the initial state (\ref{Mott}):
\be
\phi_s(0)~=~e^{i\theta_s(0)}~
\label{randomphases}
\ee
with independent random phases $\theta_s(0)$. 

TWA \cite{TW,TWAAnatoli} is a semi-classical approximation accurate for sufficiently short quench times $\tau_Q$. In the Appendix we extract the largest $\tau_Q$ where it is still applicable from simulations in small systems of a few lattice sites, and in Eq. (\ref{estimate}) below we give an instanton-based estimate for large system sizes. Both estimates predict the largest $\tau_Q$ to grow with the density $n$. 

Here we focus on the integer winding number
\be
W ~=~
\frac{1}{2\pi}\sum_{s=1}^N \left.\left(\theta_{s+1}-\theta_s\right)\right|_{\in(-\pi,\pi)} ~,
\ee
where $\theta_s={\rm arg}(\phi_s)$ and each phase step $\theta_{s+1}-\theta_s$ between nearest neighbor sites is brought to the interval $(-\pi,\pi)$ modulo $2\pi$. 

\begin{figure}[h]
\includegraphics[width=0.99\columnwidth,clip=true]{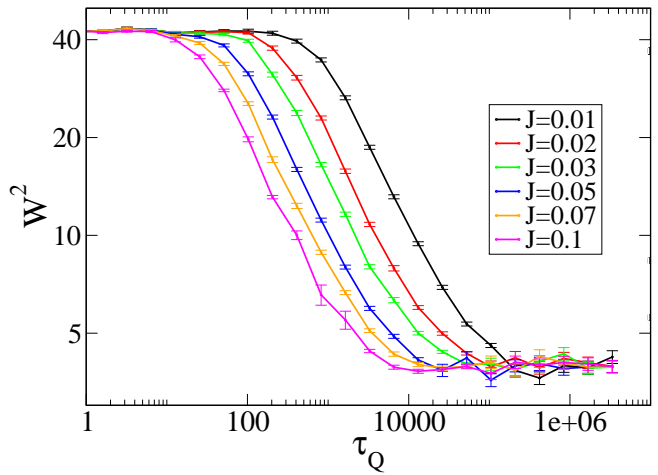}
\caption{ 
Results from simulations using truncated Wigner method (\ref{GPE},\ref{randomphases}) averaged over $2~10^4$ realizations.
The plot shows variance of the winding number $\overline{W^2}$ on $N=512$ sites as a function of $\tau_Q$. For each $J$ there is a threshold quench time $\tau_Q^c(J)$ above which the variance $\overline{W^2}$ saturates. The saturated variance does not depend on $J$. 
}
\label{FigNumericsA}
\end{figure}

Figure \ref{FigNumericsA} shows average square of the winding number $\overline{W^2}$, equal to its variance, at a final $J$ as a function of the quench time $\tau_Q$. 
One can distinguish three different regimes of $\tau_Q$:

\bi 

\item For small $\tau_Q$ there is not enough time for phases at different sites to become correlated, the final $\phi_s$ remain close to the initial fields (\ref{randomphases}), and the random initial phases result in large winding numbers of variance $\overline{W^2}=N/12$; 

\item For longer $\tau_Q$ there is more time to correlate nearest neighbor phases and, in accordance with the KZM \cite{Meisner,preMeisner,sols}, $\overline{W^2}$ decays with increasing $\tau_Q$. One might expect this KZM decay to extrapolate to $\tau_Q\to\infty$.

\item Contrary to this natural expectation, we find that for even longer $\tau_Q$ there is a crossover quench time $\tau_Q^c$ where the variance $\overline{W^2}$ saturates at a finite value which is the same for all $J$. Quite surprisingly, above $\tau_Q^c$ the winding number does not depend on $\tau_Q$. 

\ei

%%%%%%%%%%%%%%%%%%%%%%%%%%%%%%%%%%%%%%%%%%%%%%%%%%%%%%%%%%%%%%%%%%%%%%%%%%%%
%%%%%%%%%%%%%%%%%%%%%%%%%%%%%%%%%%%%%%%%%%%%%%%%%%%%%%%%%%%%%%%%%%%%%%%%%%%%

%%%%%%%%%%%%%%%%%%%%%%%%%%%%%%%%%%%%%%%%%%%%%%%%%%%%%%%%%%%%%%%%%%%%%%%%%%%%
\begin{figure}[h]
\includegraphics[width=0.99\columnwidth,clip=true]{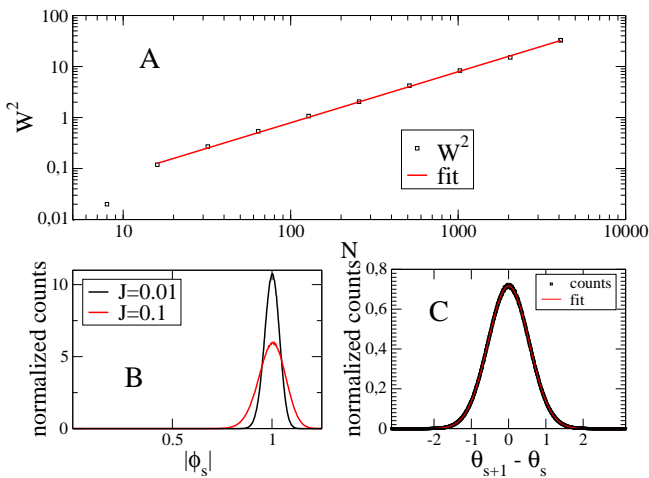}
\caption{ 
Results from simulations using truncated Wigner method (\ref{GPE},\ref{randomphases}) averaged over $2~10^4$ realizations:
In A, the saturated variance $\overline{W^2}$ at $J=0.01$ and $\tau_Q=52428.8$ as a function of lattice size $N$. The solid line is a linear fit $\overline{W^2}=0.0079N$ for $N\geq16$. At $N=8$ the variance is below the linear fit and at $N=4$ (not shown) it is zero.   
In B, histograms of the modulus $|\phi_s|$ at $J=0.01$ for $\tau_Q=52428.8$ and $N=512$. In the Josephson regime $J\ll1$ fluctuations of the modulus around $|\phi_s|=1$ are small.
In C, a generic histogram of phase steps $\theta_s-\theta_{s+1}$, here at $J=0.01$, $\tau_Q=52428.8$ and $N=512$. The solid line is a Gaussian fit. 
}
\label{FigNumericsBCD}
\end{figure}
%%%%%%%%%%%%%%%%%%%%%%%%%%%%%%%%%%%%%%%%%%%%%%%%%%%%%%%%%%%%%%%%%%%%%%%%%%%%

%%%%%%%%%%%%%%%%%%%%%%%%%%%%%%%%%%%%%%%%%%%%%%%%%%%%%%%%%%%%%%%%%%%%%%%%%%%%%%%%%%%%

\section{ Josephson equations }

%%%%%%%%%%%%%%%%%%%%%%%%%%%%%%%%%%%%%%%%%%%%%%%%%%%%%%%%%%%%%%%%%%%%%%%%%%%%%%%%%%%%%%

This unexpected result can be more readily explained in terms of the Josephson equations. We can always write $\phi_s=\left(1+f_s\right)e^{i\theta_s}$ in Eq. (\ref{GPE}) with real $f_s,\theta_s$. In the Josephson regime we have $|f_s|\ll1$, see Fig. \ref{FigNumericsBCD}B. After elimination of $f_s$ we obtain Josephson equations
\be
\frac{d^2\theta_s}{dt^2}=
2J(t)
\left[
\sin\left(\theta_{s+1}-\theta_s\right)-\sin\left(\theta_s-\theta_{s-1}\right)
\right]~.
\label{JosephsonJ}
\ee
In case of constant $J$ and more than $2$ sites these equations are chaotic.

It is convenient to eliminate $\tau_Q$ by introducing a rescaled time variable,  
\be
u~=~t~\tau_Q^{-1/3}~=~J(t)~\tau_Q^{2/3}~,
\ee 
when we obtain dimensionless chaotic equations
\be
\frac{d^2\theta_s}{du^2}=
2u
\left[
\sin\left(\theta_{s+1}-\theta_s\right)-\sin\left(\theta_s-\theta_{s-1}\right)
\right]
\label{Josephson}
\ee
with random initial phases $\theta_s(0)$ and, for large $n$, vanishing initial velocities $\frac{d\theta_s}{du}(0)=0$.

%%%%%%%%%%%%%%%%%%%%%%%%%%%%%%%%%%%%%%%%%%%%%%%%%%%%%%%%%%%%%%%%%%%%%%%%%%%%%%%%%%%%%%%%%%%%%%%%%%%%%%%%%%%%%%%%%%%%%%%%%%%%%%%%%%%%%%%%%%%%%%%%%%

\section{ Thermalization } 

%%%%%%%%%%%%%%%%%%%%%%%%%%%%%%%%%%%%%%%%%%%%%%%%%%%%%%%%%%%%%%%%%%%%%%%%%%%%%%%%%%%%%%%%%%%%%%%%%%%%%%%%%%%%%%%%%%%%%%%%%%%%%%%%%%%%%%%%%%%%%%%%%%

The (dimensionless) nonlinear system (\ref{Josephson}) approximately thermalizes after rescaled time 
$
\hat u~\simeq~ 1~
%\label{hatu}
$ 
in the sense that averages of local observables can be obtained from a Boltzmann distribution. Since the Hamiltonian is time-dependent (\ref{H},\ref{Jt}), the temperature depends on time and the thermal distribution is only approximate. Nevertheless, in first approximation the evolution after $\hat u$ can be considered an adiabatic process with the state of the system following closely to the instantaneous state of thermal equilibrium.

\begin{figure}[h]
\includegraphics[width=0.85\columnwidth,clip=true]{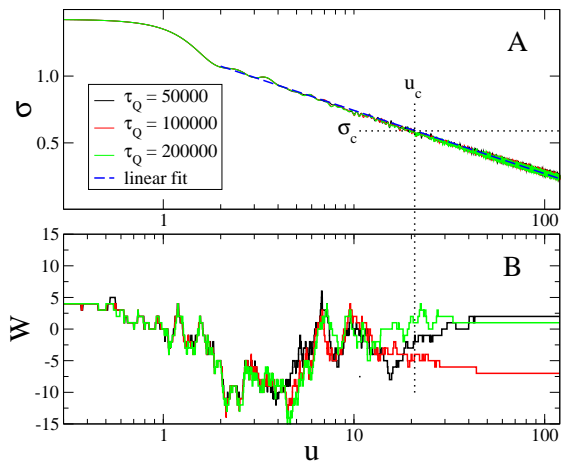}
\caption{ 
In panel A, estimated dispersion $\sigma$ of a phase step between nearest neighbor lattice 
sites as a function of $u$ on a lattice of $N=2048$ sites for three different $\tau_Q$. 
The three plots collapse confirming that the rescaled time $u$ is the relevant time variable. 
In the range $2<u<120$ the collapsed log-log plots are linear with a slope $-0.28$ 
implying $\sigma\sim u^{-0.28}$. In panel B, corresponding evolutions of the winding number.
$W$ is trapped after $\sigma$ shrinks below $\sigma_c$ at $u_c\simeq20$.
}
\label{FigJumps}
\end{figure}

After $\hat u$ the variance 
\be 
\sigma^2~=~\overline{\left(\theta_{s+1}-\theta_s\right)^2}~
\ee
of phase steps $\theta_{s+1}-\theta_s$ is shrinking, see Fig. \ref{FigJumps}A, and energy of the Josephson system (\ref{JosephsonJ}) becomes approximately quadratic  
\bea
E 
&=&
\sum_{s=1}^N
\left\{
\frac12\left(\frac{d\theta_s}{dt}\right)^2+
2J(t) \left[1-\cos\left(\theta_{s+1}-\theta_s\right)\right]
\right\}~
\nonumber\\
&\approx&
\sum_{s=1}^N
\left\{
\frac12\left(\frac{d\theta_s}{dt}\right)^2+
J(t) \left(\theta_{s+1}-\theta_s\right)^2
\right\}~.
\label{EJosephson}
\eea
Consequently, the Boltzmann distribution of phase steps $\theta_{s+1}-\theta_s$ becomes a Gaussian, see Fig. \ref{FigNumericsBCD}C, of zero mean and a variance
$\sigma^2$ related to the temperature 
\be 
T~=~2J\sigma^2
\label{T}
\ee 
of the quadratic system (\ref{EJosephson}). Here the Boltzmann constant $k_B=1$. 

Due to equipartition, thermal average of the quadratic energy (\ref{EJosephson}) is 
\be 
\langle E\rangle~=~T~N~.
\label{<E>}
\ee
On the other hand, the time-dependent $J$ in (\ref{EJosephson}) makes the thermal energy time-dependent
\be 
\frac{d}{dt}\langle E \rangle~=~\frac{dJ}{dt}~\frac{\langle E_{\rm kin}\rangle}{J}~,
\label{dEdT}
\ee
where $\langle E_{\rm kin}\rangle=\frac12TN$ is kinetic (hopping) energy, i.e., the thermal average of the last term in Eq. (\ref{EJosephson}).
A combination of Eqs. (\ref{<E>}) and (\ref{dEdT}) gives a simple differential equation
$ 
\frac{dT}{dJ}~=~\frac12\frac{T}{J}~
$
or, equivalently,
\be 
T~\propto~J^{1/2}~.
\label{TJ}
\ee
This is the time-dependent temperature in the adiabatic process after $\hat u$. 

A missing multiplicative integration constant in Eq. (\ref{TJ}) can be determined by an approximate initial condition that $\langle E \rangle\simeq \hat J N$ at $\hat u$. Here $\hat J\simeq\tau_Q^{-2/3}$ corresponds to $\hat u\simeq 1$ and the initial condition assumes that at $\hat u$ the phases are (almost) as random as in the initial Mott state. With this initial condition at $\hat u$ the temperature in the following adiabatic process becomes
\be 
T~\simeq~\hat J^{1/2}J^{1/2}~\simeq~\frac{J^{1/2}}{\tau_Q^{1/3}}~.
\ee
With Eq. (\ref{T}) this equation translates to
\be 
\sigma^2~\simeq~J^{-1/2}~\tau_Q^{-1/3}~=~u^{-1/2}~
\ee
after $\hat u\simeq1$. This scaling, in an equivalent form $\sigma\simeq u^{-1/4}$, is confirmed by numerical results in Fig. \ref{FigJumps} A.

%%%%%%%%%%%%%%%%%%%%%%%%%%%%%%%%%%%%%%%%%%%%%%%%%%%%%%%%%%%%%%%%%%%%%%%%%%%%
%%%%%%%%%%%%%%%%%%%%%%%%%%%%%%%%%%%%%%%%%%%%%%%%%%%%%%%%%%%%%%%%%%%%%%%%%%%%

%%%%%%%%%%%%%%%%%%%%%%%%%%%%%%%%%%%%%%%%%%%%%%%%%%%%%%%%%%%%%%%%%%%%%%%%%%%%%%%%%%%%

\section{ Ergodicity breaking }

%%%%%%%%%%%%%%%%%%%%%%%%%%%%%%%%%%%%%%%%%%%%%%%%%%%%%%%%%%%%%%%%%%%%%%%%%%%%%%%%%%%%%%

\begin{figure}[t]
\includegraphics[width=0.8\columnwidth,clip=true]{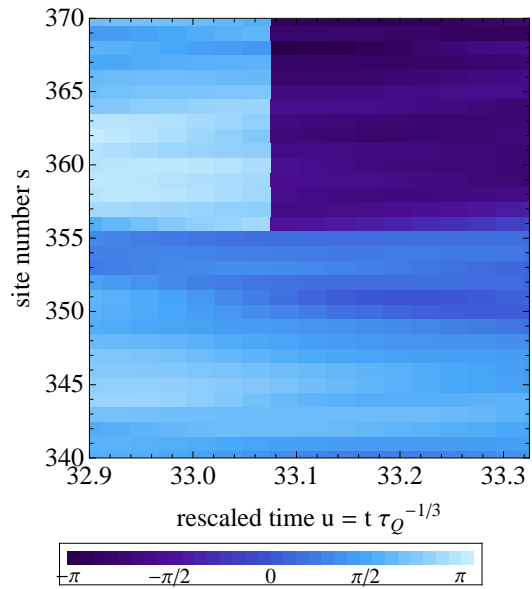}
\caption{ Integrated phase
$
\Theta_s=\sum_{j=1}^s \left.\left(\theta_{j+1}-\theta_j\right)\right|_{\in(-\pi,\pi)}
$
as a function of the site number $s$ and the rescaled time $u$. Notice that $W=\Theta_N/2\pi$. At $u\approx33.07$ the winding number jumps by $-1$ at the link between sites $355$ and $356$, i.e., the phase step $\left.(\theta_{356}-\theta_{355})\right|_{\in(-\pi,\pi)}$ jumps from $+\pi$ to $-\pi$ across the saddle separating different $W$.
}
\label{FigSmoking}
\end{figure}

The accuracy of the quadratic approximation in (\ref{EJosephson}) suggests that with shrinking $\sigma^2$ the system crosses over from ergodic to increasingly regular behaviour.  
The most striking manifestation of the increasing regularity is ergodicity breaking between different integer values of the winding number $W$, see Fig. \ref{FigJumps}B. In energy landscape picture, valleys with different integer $W$ are separated by an unstable saddle-point ``phase slip'' solution. In the Josephson regime $J\ll1$ the phase slip is localized on a single link with a phase step $\theta_{s+1}-\theta_{s}=\pm\pi$. The frequent initial jumps of $W$ shown in Fig. \ref{FigJumps}B all pass through the localized phase slip, see an example in Fig. \ref{FigSmoking}.

%%%%%%%%%%%%%%%%%%%%%%%%%%%%%%%%%%%%%%%%%%%%%%%%%%%%%%%%%%%%%%%%%%%%%%%%%%%%
%%%%%%%%%%%%%%%%%%%%%%%%%%%%%%%%%%%%%%%%%%%%%%%%%%%%%%%%%%%%%%%%%%%%%%%%%%%%

According to the LAMH theory \cite{LAMH,instanton}, when $|W/N|\ll1$ the frequency of the winding number jumps is $\propto e^{-\beta 4J}$, where the $4J$ is energy (\ref{EJosephson}) of the localized phase slip. Since the temperature is $\beta^{-1}=2J\sigma^2$ in Eq. (\ref{T}), this activation coefficient is $\propto e^{-2/\sigma^2}$ and the integer winding number freezes out when $\sigma$ falls below 
\be 
\sigma_c~\simeq~1~.
\ee
From the data in Fig. \ref{FigJumps}B this happens at time $u_c\simeq 20-50$. This is when ergodicity between different $W$'s breaks down and the winding number (distribution) gets stuck.  

%%%%%%%%%%%%%%%%%%%%%%%%%%%%%%%%%%%%%%%%%%%%%%%%%%%%%%%%%%%%%%%%%%%%%%%%%%%%%%%%%%%%

\section{ Trapped winding number }

%%%%%%%%%%%%%%%%%%%%%%%%%%%%%%%%%%%%%%%%%%%%%%%%%%%%%%%%%%%%%%%%%%%%%%%%%%%%%%%%%%%%%%

Below $\sigma_c$ the phase step dispersion $\sigma$ keeps shrinking, so the random walk $\theta_s(u)$ keeps smoothing, but its temperature is not enough to induce jumps unwinding its net winding number. Thus the frozen winding number is a remnant of a random walk with $\sigma=\sigma_c$ and its variance is
\be
\overline{W^2}~\simeq~\frac{1}{(2\pi)^2}~N~\sigma_c^2~.
\label{W2}
\ee 
Here $N\sigma_c^2$ is average distance-squared ``random walked'' by the phase around the ring. The estimate (\ref{W2}) agrees with the linear fit $\overline{W^2}=0.0079N$ in Fig. \ref{FigNumericsBCD}A when
\be 
\sigma_c~\approx~0.56~.
\label{sigmac}
\ee
For large $N$ the linear scaling (\ref{W2}) gives stronger winding than typical winding originating from quantum fluctuations in the ground state that is only logarithmic in $N$. 

The winding number assumes a fixed value at a rescaled time $u_c$ when $\sigma$ falls below $\sigma_c$. For a given final $J$, as in Fig. \ref{FigNumericsA}, $u_c$ translates to a quench time 
\be
\tau_Q^c ~\simeq~ u_c^{3/2}J^{-3/2} ~
\label{tauQc}
\ee
which is independent of the lattice size $N$ and whose dependence on $J$ is consistent with Fig. \ref{FigNumericsA}. When $\tau_Q\gg\tau_Q^c$ the variance $\overline{W^2}$ saturates at the finite value in Eq. (\ref{W2}). $W$ settles down at $J_c\simeq u_c\tau_Q^{-2/3}$ which is in the Josephson regime, $J_c\ll1$, for slow enough quenches with $\tau_Q\gg u_c^{3/2}$. 

On the other hand, for $\tau_Q\ll u_c^{3/2}\approx 10^3$, the familiar KZM scaling $\overline{W^2}\sim \tau_Q^{-1/3}$ was observed \cite{Meisner} in quenches that take the system beyond the Josephson regime into $J\gg 1$ territory. 

After the thermal freeze-out at $J_c$ the winding number could still change by quantum tunneling. Its rate $\Gamma\propto e^{-a n\sqrt{J_c}}$ with a numerical constant $a\simeq1$ can be obtained from instanton calculations \cite{instanton}. It is negligible when $n\sqrt{J_c}\gg1$ or, equivalently, 
\be 
\tau_Q~\ll~n^3~.
\label{estimate}
\ee
With, say, $n=100$ particles per site this is a generous upper estimate for the range of $\tau_Q$ when the semiclassical TWA is applicable. The same estimate (\ref{estimate}) is also obtained from the condition that the winding number freezes out at a value of tunneling rate far above the Mott transition: $J_c\gg n^{-2}$. Indeed, near the Mott transition, where the discrete nature of site occupation numbers is essential, the semiclassical TWA is not applicable.

%%%%%%%%%%%%%%%%%%%%%%%%%%%%%%%%%%%%%%%%%%%%%%%%%%%%%%%%%%%%%%%%%%%%%%%%%%%%%%%%%%%%

\section{ Phonons }\label{SectionPhonons}

%%%%%%%%%%%%%%%%%%%%%%%%%%%%%%%%%%%%%%%%%%%%%%%%%%%%%%%%%%%%%%%%%%%%%%%%%%%%%%%%%%%%%%

After the winding number freezes out, it is convenient to think about a smooth persistent flow with phononic fluctuations on top of it. Pushing $\tau_Q$ beyond $\tau_Q^c$ (or the rescaled time $u$ beyond the freezing time $u_c$) makes no difference for the frozen winding number $W$, but it does make a difference for the phonons. Soon after the freeze-out most energy is deposited in the phononic fluctuations, but as $\sigma$ keeps shrinking below $\sigma_c$, the field tends to a smooth persistent flow
\be
\phi_s~=~e^{2\pi iWs/N}~,
\label{phiW}
\ee 
where $W$ is the winding number frozen near $\sigma_c$ and the circulation has constant phase steps $\theta_{s+1}-\theta_s=2\pi W/N$. 

The regular behaviour in this regime can be described by small phase fluctuations $\psi_s$ (phonons) around this smooth background $\theta_s=2\pi Ws/N$ satisfying a linearized version of the Josephson equations (\ref{Josephson}). Indeed, their linearization in the small phase fluctuations yields
\be
\frac{d^2\psi_s}{du^2}~=~
c~u
\left(
\psi_{s+1}-2\psi_s+\psi_{s-1}  
\right)~,
\label{Hermitean}
\ee
where $c=2\cos(2\pi W/N)$. In pseudomomentum representation $\psi_s=\sum_k\alpha_k\exp(iks)$, where $\alpha_k$ satisfy the Airy equations 
\be 
\frac{d^2\alpha_k}{du^2}~=~-2u~c~(1-\cos k)~\alpha_k~.
\ee
Since for large $u$ the envelopes of Airy functions decay like $\alpha_k\sim u^{-1/4}$, then the average square of a phase step between nearest neighbor sites is 
\bea
\sigma^2 &=&
\overline{\left(\theta_{s+1}+\psi_{s+1}-\theta_s-\psi_s\right)^2}
\nonumber\\
&=&
\overline{\left( \frac{2\pi W}{N} + \psi_{s+1}-\psi_s \right)^2}
\nonumber\\
&=&
\left(\frac{2\pi}{N}\right)^2\overline{W^2}+\overline{\left(\psi_{s+1}-\psi_s\right)^2}
\nonumber\\
&=&
\frac{\sigma_c^2}{N}+\frac{A}{u^{1/2}}
\label{A}~.
\eea
Here we used Eq. (\ref{W2}) and assumed zero correlation between $W$ and small phase fluctuations $\psi_{s+1}-\psi_s$. 

\begin{figure}[h]
\includegraphics[width=0.8\columnwidth,clip=true]{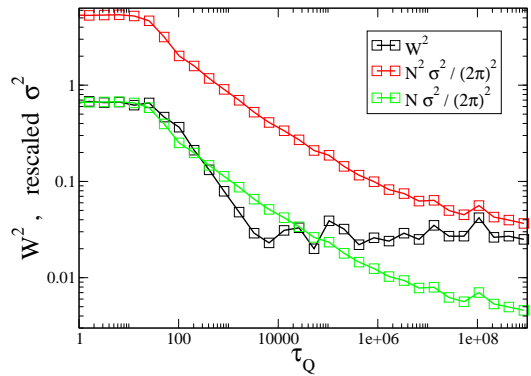}
\caption{ 
$\overline{W^2}$ and two rescaled $\sigma^2$ on $N=8$ sites at $J=0.1$. For small $\tau_Q$ 
the winding number $W$ originates from a random walk of phase around the lattice and, consequently, 
$\overline{W^2}\approx N\sigma^2/(2\pi)^2$. In contrast, for large $\tau_Q$ there is an ensemble 
of smooth fields $\phi_s=\exp\left(2\pi iWs/N\right)$ with a random $W$ and, consequently, 
$\sigma^2\approx\overline{(2\pi W/N)^2}$ equivalent to $\overline{W^2}\approx N^2\sigma^2/(2\pi)^2$.
}
\label{FigAdiabatic}
\end{figure}

The unknown constant $A\simeq\sigma_c^2u_c^{1/2}$ in Eq. (\ref{A}) that measures the overall magnitude of phase fluctuations $\psi_s$ can be fixed by the initial condition $\sigma^2(u_c)\simeq\sigma_c^2$ when the winding number is freezing. In this way we obtain scaling
\be
\sigma^2\left(u\ll N^2u_c\right) ~\sim~ u^{-1/2}~
\ee
soon after the freeze-out of the winding number. This scaling is confirmed in Fig. \ref{FigJumps}A. On the other hand, for late enough times we find a saturation at
\be
\sigma^2\left(u\gg N^2u_c\right) ~=~ 
\frac{\sigma^2_c}{N}~ = ~ 
\frac{(2\pi)^2}{N^2}~\overline{W^2}~
\ee
which is confirmed in Fig. \ref{FigAdiabatic}. This saturation means that asymptotically the phase step $\theta_{s+1}-\theta_s$ is dominated by the smooth winding number $2\pi W/N$ with negligible phase fluctuations $\psi_s$.   

Finally, in agreement with Ref. \cite{Sorin}, the persistent flow (\ref{phiW}) in the Josephson regime is not stable for $N\leq4$ when $c=2\cos(2\pi W/N)\leq0$ in Eq. (\ref{Hermitean}). This explains why $\overline{W^2}=0$ for $N=4$ and it is below the large-$N$ linear fit for $N=8$, see Fig. \ref{FigNumericsBCD}A.

%%%%%%%%%%%%%%%%%%%%%%%%%%%%%%%%%%%%%%%%%%%%%%%%%%%%%%%%%%%%%%%%%%%%%%%%%%%%
%%%%%%%%%%%%%%%%%%%%%%%%%%%%%%%%%%%%%%%%%%%%%%%%%%%%%%%%%%%%%%%%%%%%%%%%%%%%

%%%%%%%%%%%%%%%%%%%%%%%%%%%%%%%%%%%%%%%%%%%%%%%%%%%%%%%%%%%%%%%%%%%%%%%%%%

\section{ Conclusion } \label{SectionConclusion}

%%%%%%%%%%%%%%%%%%%%%%%%%%%%%%%%%%%%%%%%%%%%%%%%%%%%%%%%%%%%%%%%%%%%%%%%%%%%%%%%%%%%%%

A ring of isolated Bose-Einstein condensates becomes increasingly
correlated as the tunneling rate between the condensates increases. Consequently, the initial 
random walk of phase around the ring smoothes and the variance of its winding number decreases. 
However, when the phase becomes smooth enough, so that its small scale variations are no longer sufficient to let it occasionally hop across
barriers separating different integer winding numbers, the winding number freezes out.
As it is the critical smoothness that determines the trapped winding number, its variance 
does not depend on the quench time. Both this result and the underlying process are valid in a regime that allows for the ``memory loss'' and thus differs from
the paradigmatic Kibble-Zurek mechanism.

%%%%%%%%%%%%%%%%%%%%%%%%%%%%%%%%%%%%%%%%%%%%%%%%%%%%%%%%%%%%%%%%%%%%%%%%%% 
{\bf Acknowledgements. ---} We were supported by Polish Government research 
projects N202 079135 and N202 124736 (J.D. and M.T.) and by DoE under the LDRD grant at the Los Alamos National Laboratory (W.H.Z.).
%%%%%%%%%%%%%%%%%%%%%%%%%%%%%%%%%%%%%%%%%%%%%%%%%%%%%%%%%%%%%%%%%%%%%%%%%%

\vspace{1cm}

\appendix{\bf Appendix: TWA versus exact simulation in a small system }

We use TWA (i.e., in effect take the limit of infinite average density $n\to\infty$) before we investigate large $\tau_Q$. ``Mathematically speaking", with this order of limits, TWA is accurate and the system gets excited from its instantaneous ground state for any $\tau_Q$. However, this formal statement is not satisfactory for a physicist. After all, we aim to investigate large but finite systems over large but finite timescales. So, what is of interest, is to determine -- given a density (or a total number of atoms per site) $n$ -- what is a $\tau_Q(n)$ when TWA breaks down. To this end, we have carried out a supplementary numerical study that, we believe, settles this issue. 

%%%%%%%%%%%%%%%%%%%%%%%%%%%%%%%%%%%%%%%%%%%%%%%%%%%%%%%%%%%%%%%%%%%%%%%%%%%%
\begin{figure}[h!]
\includegraphics[width=0.99\columnwidth,clip=true]{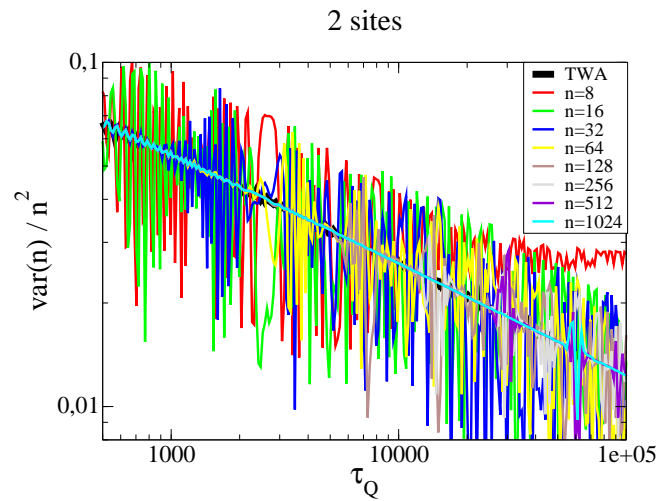}
\caption{ 
Relative number variance at 2 sites for $n=8,...,1024$.
}
\label{2sites}
\end{figure}
%%%%%%%%%%%%%%%%%%%%%%%%%%%%%%%%%%%%%%%%%%%%%%%%%%%%%%%%%%%%%%%%%%%%%%%%%%%%
%%%%%%%%%%%%%%%%%%%%%%%%%%%%%%%%%%%%%%%%%%%%%%%%%%%%%%%%%%%%%%%%%%%%%%%%%%%%
\begin{figure}[h!]
\includegraphics[width=0.99\columnwidth,clip=true]{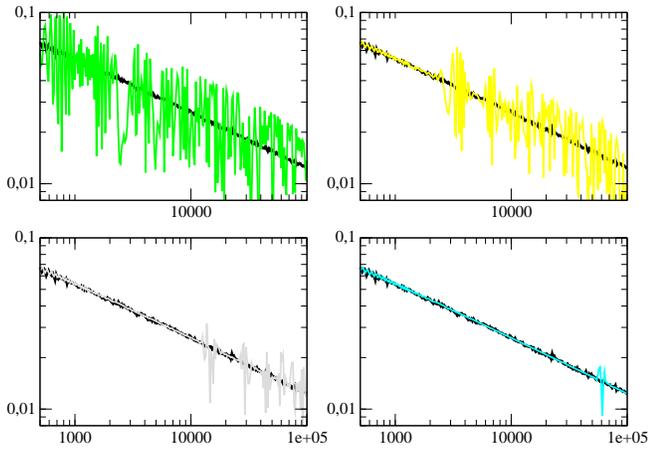}
\caption{ 
The same data as in Figure \ref{2sites}.
Relative number variance at 2 sites for $n=16,64,256,1024$, see the legend in Fig. \ref{2sites}.
}
\label{2sitesbis}
\end{figure}
%%%%%%%%%%%%%%%%%%%%%%%%%%%%%%%%%%%%%%%%%%%%%%%%%%%%%%%%%%%%%%%%%%%%%%%%%%%%
%%%%%%%%%%%%%%%%%%%%%%%%%%%%%%%%%%%%%%%%%%%%%%%%%%%%%%%%%%%%%%%%%%%%%%%%%%%%
\begin{figure}[h!]
\includegraphics[width=0.99\columnwidth,clip=true]{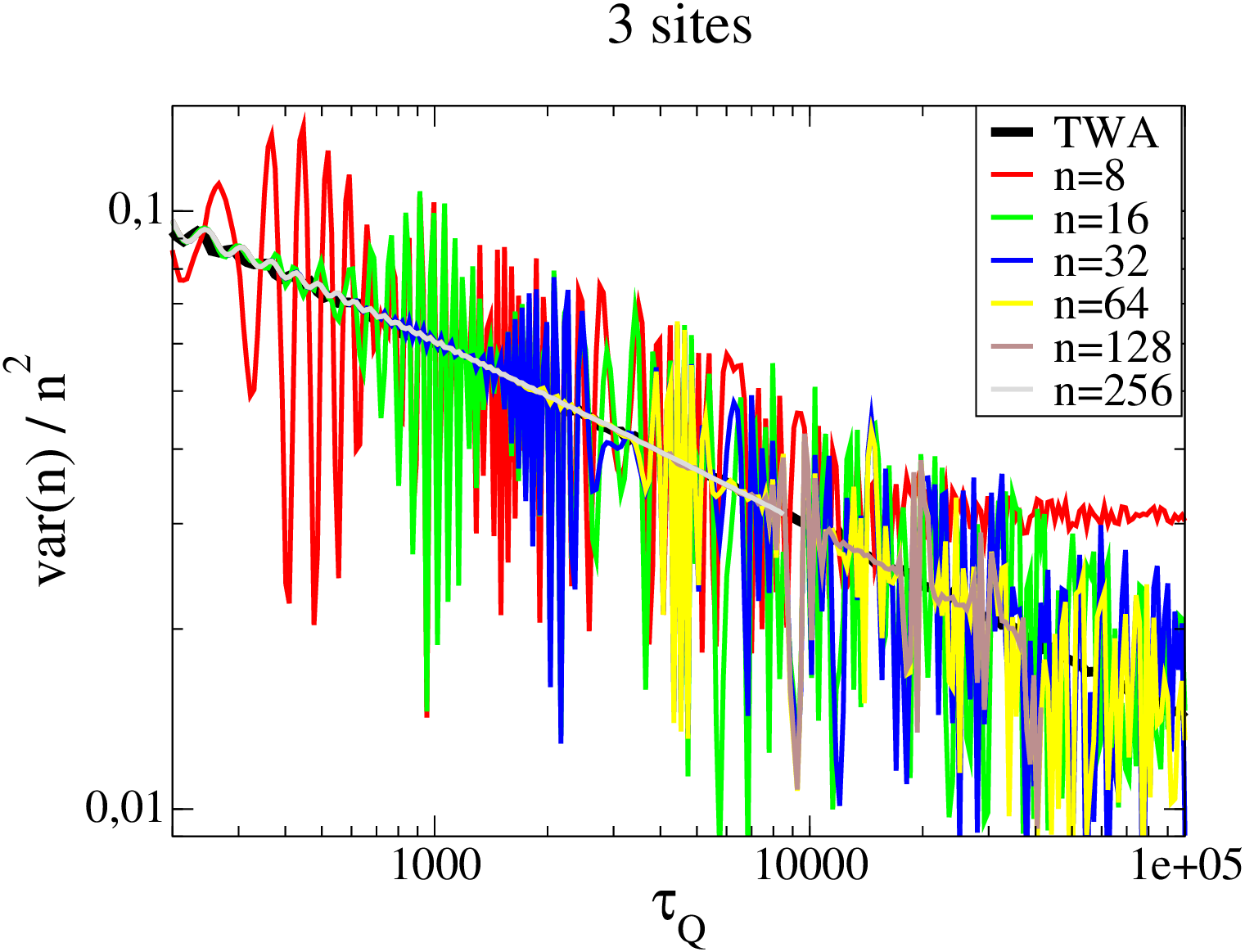}
\caption{ 
Relative number variance at 3 sites.
}
\label{3sites}
\end{figure}
%%%%%%%%%%%%%%%%%%%%%%%%%%%%%%%%%%%%%%%%%%%%%%%%%%%%%%%%%%%%%%%%%%%%%%%%%%%%
%%%%%%%%%%%%%%%%%%%%%%%%%%%%%%%%%%%%%%%%%%%%%%%%%%%%%%%%%%%%%%%%%%%%%%%%%%%%
\begin{figure}[h!]
\includegraphics[width=0.99\columnwidth,clip=true]{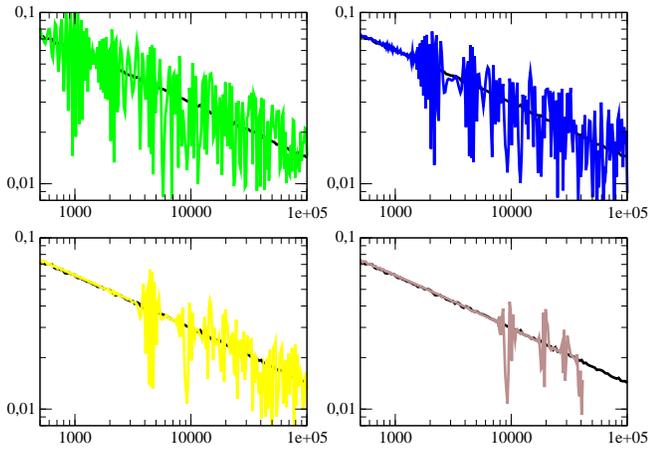}
\caption{ 
The same data as in Figure \ref{3sites}.
Relative number variance at 3 sites for $n=16,32,64,128$, see the legend in Fig. \ref{3sites}.
}
\label{3sitesbis}
\end{figure}
%%%%%%%%%%%%%%%%%%%%%%%%%%%%%%%%%%%%%%%%%%%%%%%%%%%%%%%%%%%%%%%%%%%%%%%%%%%%
%%%%%%%%%%%%%%%%%%%%%%%%%%%%%%%%%%%%%%%%%%%%%%%%%%%%%%%%%%%%%%%%%%%%%%%%%%%%
\begin{figure}[h!]
\includegraphics[width=0.99\columnwidth,clip=true]{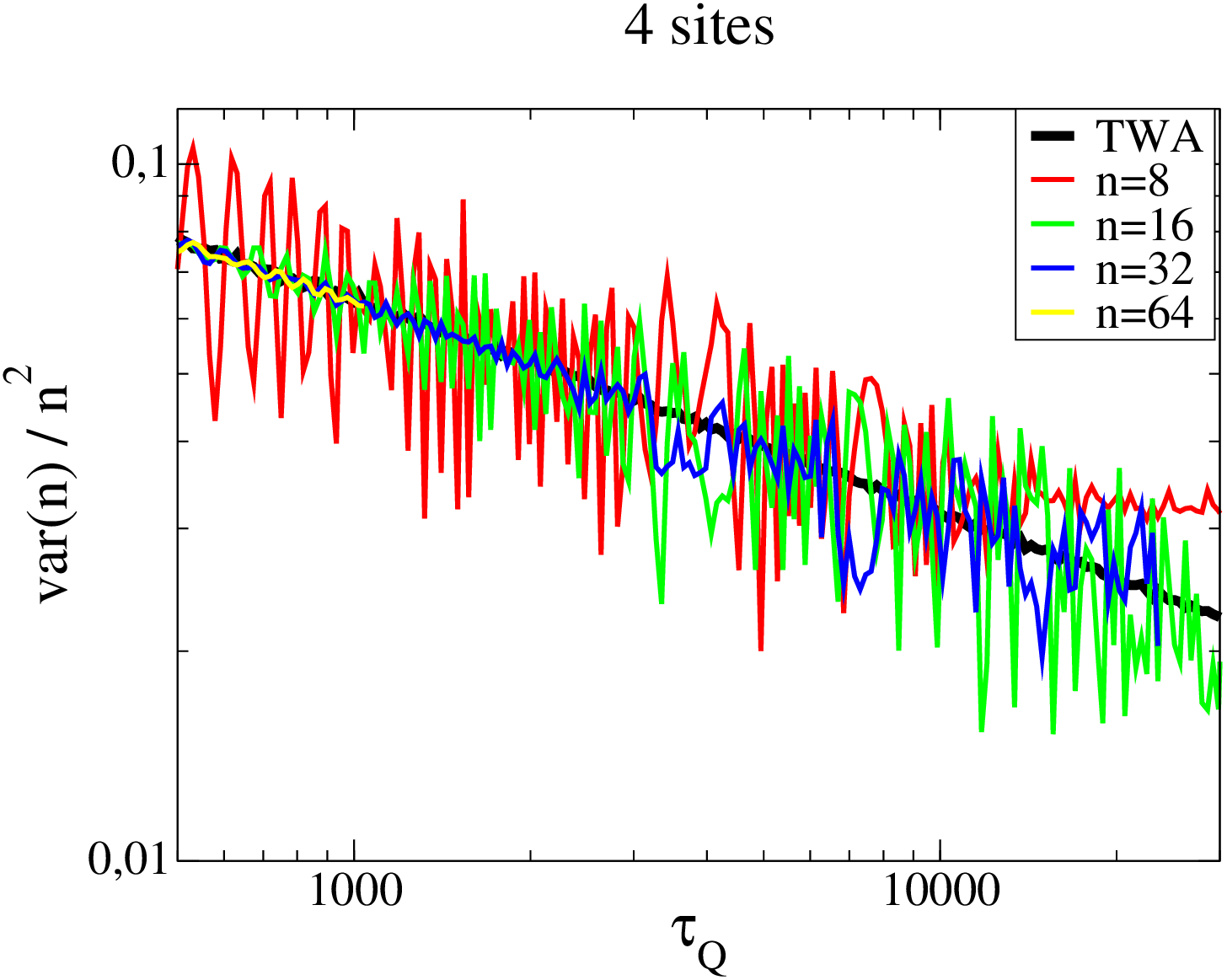}
\caption{ 
Relative number variance at 4 sites.
}
\label{4sites}
\end{figure}
%%%%%%%%%%%%%%%%%%%%%%%%%%%%%%%%%%%%%%%%%%%%%%%%%%%%%%%%%%%%%%%%%%%%%%%%%%%%
%%%%%%%%%%%%%%%%%%%%%%%%%%%%%%%%%%%%%%%%%%%%%%%%%%%%%%%%%%%%%%%%%%%%%%%%%%%%
\begin{figure}[h!]
\includegraphics[width=0.99\columnwidth,clip=true]{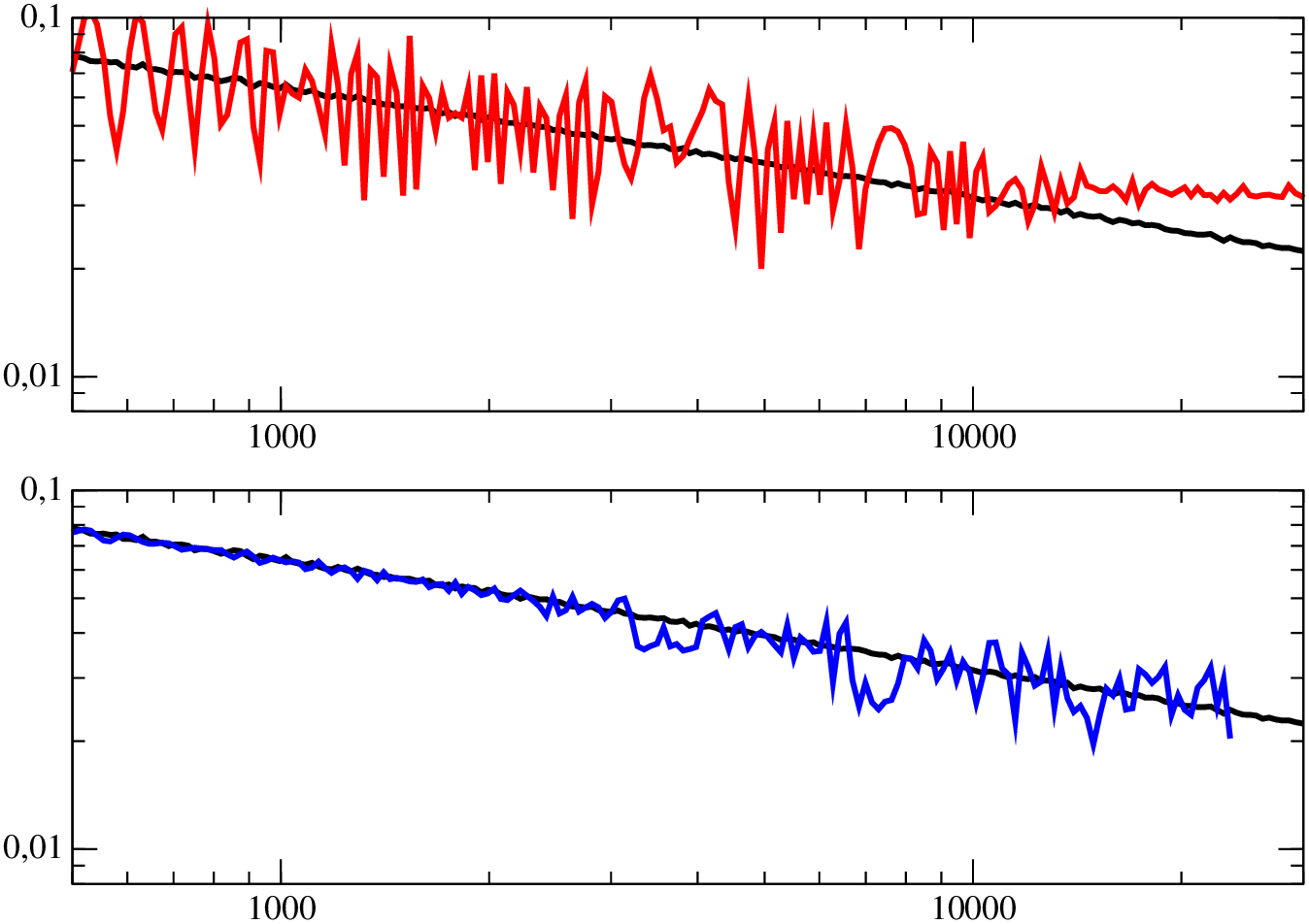}
\caption{ 
The same data as in Figure \ref{4sites}.
Relative number variance at 4 sites for $n=8,32$, see the legend in Fig. \ref{4sites}.
}
\label{4sitesbis}
\end{figure}
%%%%%%%%%%%%%%%%%%%%%%%%%%%%%%%%%%%%%%%%%%%%%%%%%%%%%%%%%%%%%%%%%%%%%%%%%%%%

\begin{figure}[h!]
\includegraphics[width=0.99\columnwidth,clip=true]{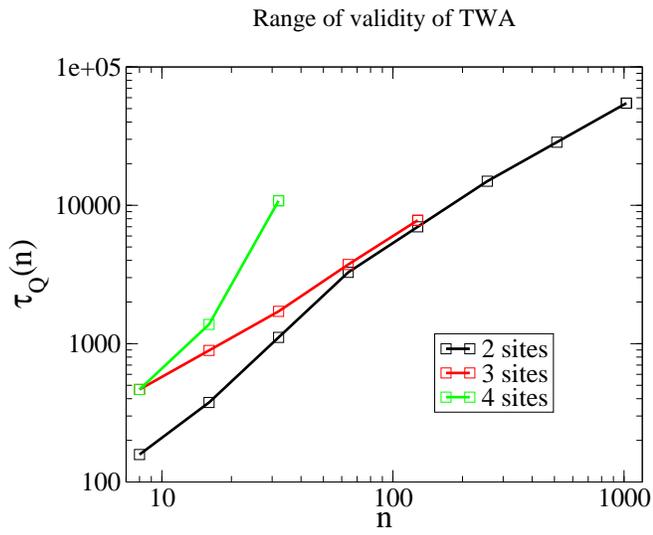}
\caption{ 
The largest range of accuracy of TWA $\tau_Q(n)$ as a function of density $n$ for a system size of $2,3,4$ sites. Here we define $\tau_Q(n)$ as the $\tau_Q$ when the exact variance deviates from the TWA by more than $20\%$ for the first time.
}
\label{tauQn}
\end{figure}

Pairs of Figures (\ref{2sites},\ref{2sitesbis}), (\ref{3sites},\ref{3sitesbis}), and (\ref{4sites},\ref{4sitesbis}) present results of exact simulations with $n$ up to $1024$ on $2$, $3$, and $4$ sites respectively. They show relative variance of an occupation number $\hat n_s=a_s^\dag a_s$,
\be
\frac{{\rm var}(\hat n_s)}{n^2},
\ee
as a function of $\tau_Q$. The same figures show results from TWA (black solid line) averaged over $10^4$ realisations. The exact results follow the TWA up to $\tau_Q(n)$ that increases with $n$, see Figure \ref{tauQn}. The five right-most data points on $2$ sites can be fitted with
\be
\tau_Q(n)=49.5~ n^{1.01}
\label{lin}
\ee
with error bars on the last digits. A similar fit to all five data points on $3$ sites yields $\tau_Q(n)=52.0~ n^{1.02}$.
Within the error bars these are linear fits. Generally, both increasing $n$ and increasing number of sites (total number of particles) extend the range of validity of TWA. 

The linear fit (\ref{lin}) can be explained as follows. In the system of units used throughout our paper the interaction strength in the Hamiltonian (\ref{H}) is 
\be
U=\frac{1}{n}~.
\ee
The case of $2$ sites was discussed and tested by numerical simulations in Ref. \cite{TWAAnatoli}. TWA was found exact for simulations times limited by $t\ll J/U$, where $J$ was a constant tunneling rate. Extrapolating this result to our $U=1/n$ and time-dependent $J=t/\tau_Q$ this inequality becomes 
$$
\tau_Q~\ll~n~
$$
and TWA should break down above $\tau_Q(n)\sim n$, in agreement with the linear fit (\ref{lin}). 
%In addition to this, the data collected in Fig. \ref{tauQn} show a tendency of $\tau_Q(n)$ to increase with the number of sites. 

%%%%%%%%%%%%%%%%%%%%%%%%%%%%%%%%%%%%%%%%%%%%%%%%%%%%%%%%%%%%%%%%%%%%%%%%%%%%
%%%%%%%%%%%%%%%%%%%%%%%%%%%%%%%%%%%%%%%%%%%%%%%%%%%%%%%%%%%%%%%%%%%%%%%%%%%%


\begin{thebibliography}{99}

\bibitem{QuantumKZ} B. Damski, 
                    Phys. Rev. Lett. {\bf 95}, 035701 (2005);
                    W.H. Zurek {\it et al.},
                    Phys. Rev. Lett. {\bf 95}, 105701 (2005);
                    J. Dziarmaga, 
                    Phys. Rev. Lett. {\bf 95}, 245701 (2005);
                    A. Polkovnikov, 
                    Phys. Rev. B {\bf 72}, R161201 (2005).

\bibitem{review} J. Dziarmaga,
                 Adv. in Phys. {\bf 59}, 1063 (2010);
                 A. Polkovnikov {\it et al.},
                 arXiv:1007.5331.

\bibitem{KZ} T.W.B. Kibble, 
             J. Phys. A {\bf 9}, 1387 (1976); 
             Phys. Rep. {\bf 67}, 183 (1980);
             Physics Today, 60, 47 (2007);
             W.H. Zurek, 
             Nature {\bf 317}, 505 (1985);
             Acta Phys. Polon. B {\bf 24}, 1301 (1993);
             Phys. Rep. {\bf 276}, 177 (1996).

\bibitem{ferro} L. E. Sadler {\it et al.}, 
                Nature (London) 443, 312 (2006).

\bibitem{Mercedes} D. R. Scherer {\it et al.},  
                   Phys. Rev. Lett. {\bf 98}, 110402 (2007);
                   R. Carretero-Gonzalez {\it et al.},  
                   Phys. Rev. A {\bf 77}, 033625 (2008).

\bibitem{Boshier} K. Henderson, {\it et al.},  
                  New J. Phys. 11 (2009) 043030.   
                 
\bibitem{TW} K. Goral {\it et al.},
             Opt. Express {\bf 8}, 92 (2001);
             M. J. Davis and S. A. Morgan,
             Phys. Rev. A {\bf 68}, 053615 (2003);
             A. Sinatra {\it et al.},
             Phys. Rev. Lett. {\bf 87}, 210404 (2001);
             J. Phys. B {\bf 35}, 3599 (2002);
             P.B. Blakie {\it et al.},  
             Adv. Phys. {\bf 57}, 363 (2008);
             A. P. Itin and P. T\"orm\"a,
             Phys. Rev. A 79, 055602 (2009).
             
\bibitem{TWAAnatoli} A. Polkovnikov,
                     Phys. Rev. A {\bf 68}, 053604 (2003);
                     Ann. Phys. {\bf 325}, 1790 (2010).               

\bibitem{KT} T.D. K\"uhner {\it et al.}, 
             Phys. Rev. B {\bf 61}, 12474 (2000); 
             B. Damski and J. Zakrzewski,  
             Phys. Rev. A {\bf 74}, 043609 (2006).

\bibitem{Kasevich} C. Orzel {\it et al.},  
                   Science {\bf 291}, 2386 (2001); 
                   A. K. Tuchman {\it et al.},  
                   Phys. Rev. A {\bf 74}, 051601 (2006);
                   W. Li {\it et al.},  
                   Phys. Rev. Lett. {\bf 98}, 040402 (2007).                 

\bibitem{Greiner} M. Greiner {\it et al.}, 
                  Nature {\bf 415}, 39 (2002);
                  Nature {\bf 419}, 51 (2002).

\bibitem{Meisner} J. Dziarmaga {\it et al.},
                  Phys. Rev. Lett. {\bf 101}, 115701 (2008).

\bibitem{preMeisner} J. Dziarmaga {\it et al.},
                     Phys. Rev. Lett. {\bf 88}, 167001 (2002);
                     F. Cucchietti {\it et al.},  
                     Phys. Rev. A {\bf 75}, 023603 (2007);

\bibitem{sols} P. Ghosh and F. Sols, 
               Phys. Rev. A {\bf 77}, 033609 (2008);
               P. Naves and R. Sch\"utzhold,
               arXiv:1008.1548; 
               C. Trefzger and K. Sengupta,
               arXiv:1008.1285.
               
\bibitem{LAMH} J. S. Langer and V. Ambegaokar, Phys. Rev. {\bf 164}, 498 (1967);
               D. E. McCumber and B. I. Halperin, Phys. Rev. B {\bf 1}, 1054 (1970).  
               
\bibitem{instanton} A. Polkovnikov, E. Altman, E. Demler, B. Halperin, and M. D. Lukin,
                    Phys. Rev. A {\bf 71}, 063613 (2005).                                 

\bibitem{Sorin} Gh.-S. Paraoanu,
                Phys. Rev. A {\bf 67}, 023607 (2003).      
                
\end{thebibliography}
\end{document}